\documentclass[aps,prc,onecolumn,superscriptaddress,preprintnumbers,showpacs,nofootinbib,floatfix]{revtex4}
\usepackage[dvips]{graphics,graphicx}
\usepackage{amsmath}
\begin{document}
\preprint{WU-HEP-01-4}
\preprint{TWC-01-5}
\title{Hydrodynamical analysis of hadronic spectra in the 130 GeV/nucleon Au+Au
collisions}
\author{Tetsufumi Hirano}
\affiliation{Physics Department, University of Tokyo, Tokyo 113-0033, Japan}
\author{Kenji Morita}
\email{morita@hep.phys.waseda.ac.jp}
\affiliation{Department of Physics, Waseda University, Tokyo 169-8555, Japan}
\author{Shin Muroya}
\affiliation{Tokuyama Women's College, Tokuyama, Yamaguchi 745-8511, Japan}
\author{Chiho Nonaka}
\affiliation{IMC, Hiroshima University, Higashi-Hiroshima,
Hiroshima, 739-8521, Japan}


\begin{abstract}
 We study one-particle spectra and a two-particle correlation function in the
 130 GeV/nucleon Au+Au collisions at RHIC by making use of a hydrodynamical
 model. We calculate the one-particle hadronic spectra and present the first
 analysis of Bose-Einstein correlation functions based on the numerical
 solution of the hydrodynamical equations which takes both longitudinal and
 transverse expansion into account appropriately. The hydrodynamical model
 provides excellent agreement with the experimental data in the
 pseudorapidity and the transverse momentum spectra of charged hadrons, the
 rapidity dependence of anti-proton to proton ratio, and almost consistent
 result for the pion Bose-Einstein correlation functions. Our numerical
 solution with simple freeze-out picture suggests the formation of the
 quark-gluon plasma with large volume and low net-baryon density.
\end{abstract}
\maketitle

Relativistic heavy ion collisions are very attracting problems which provide
us the nature of hot and dense hadronic matter \cite{QM2001}. Creation of
a new state of the matter, the quark-gluon plasma (QGP), and many kinds of
new phenomena are expected to be found in the Relativistic Heavy Ion Collider
(RHIC) experiments at BNL of which the collision energy is much higher than
any other accelerator. However, the complicated processes during the many-body
interactions and multiparticle productions are quite hard to catch clear.
Therefore, a simple phenomenological description is indispensable for the
better understanding of the phenomena. The aims of this paper are, based on a
hydrodynamical model, to draw a simple and clear picture of the space-time
evolution of the hot and dense matter produced in the high energy heavy ion
collisions at RHIC and to give a possible explanation for the recent
experimental results.

We use a (3+1)-dimensional hydrodynamical model \cite{Ishii_PRD46} to
describe the space-time evolution assuming the local thermal and chemical
equilibrium. Several authors have already discussed RHIC results based on
hydrodynamical models. \citeauthor{Kolb_PLB500} \cite{Kolb_PLB500} discussed
anisotropic flow by making use of a (2+1)-dimensional hydrodynamic model in
which Bjorken's ansatz \cite{Bjorken} was used for the beam direction. Hence,
their discussion is limited only in the midrapidity region. 
\citeauthor{Zschiesche_nucl7037} \cite{Zschiesche_nucl7037}
discussed HBT radii based on a hydrodynamical model with use of the Bjorken's
scaling solution in the collision direction. 
One of the authors (T.H.) has already reproduced the both the pseudorapidity
and the transverse momentum spectra of hadrons by using a full
(3+1)-dimensional hydrodynamical model in Ref.\ \cite{Hirano_PRC}, where
main theme of the analysis is also anisotropic flow. 
In this paper, we focus our discussion on \textit{central}
collisions by assuming the cylindrical symmetry of the system. We calculate
the one-particle hadronic spectra and present the first analysis of
Bose-Einstein correlation functions based on the numerical solution of the
hydrodynamical equations which takes both longitudinal and transverse
expansion into account appropriately \cite{Ishii_PRD46}.

\begin{figure}[ht]
  \includegraphics[width=3.375in]{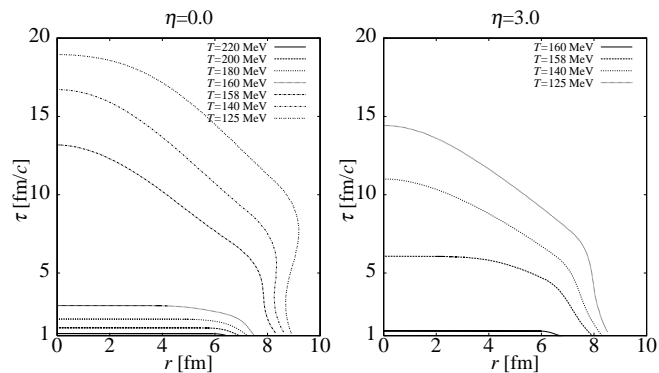}
 \caption{\label{fig:evo}Temperature contour plot on $r$-$\tau$ plane. Left
 and right figures stand for $\eta=0$ and $\eta=3$, respectively.}
\end{figure}

The hydrodynamical equations are given as $\partial_\mu T^{\mu\nu}(x) =0$
with the baryon number conservation law 
$\partial_\mu n_{\text{B}}^\mu(x) =0$.
We numerically solve these coupled equations for the perfect fluid by
the method described in Ref.\ \cite{Ishii_PRD46}. Our numerical solution
keeps entropy, energy and net baryon number conserved within 5\% of accuracy
throughout the calculation with the time step $\delta\tau=0.01$ fm/$c$. 
As for an equation of state
(EOS), we adopt a bag model EOS in which phase transition of first order
takes place \cite{Nonaka_EPJC}. The QGP phase is a free gas with a bag
constant $B$. The gas consists of massless quarks of three flavor and
gluons. The hadronic phase is a free resonance gas with an excluded
volume correction in which all resonances up to 2 GeV/$c^2$ of mass are
included. These two phases are connected by the condition of pressure
continuity \cite{Nonaka_EPJC}. 
\begin{figure}[ht]
  \includegraphics[width=3.375in]{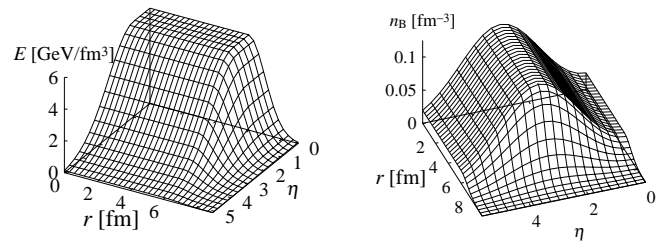}
 \caption{\label{fig:initial}Initial energy density (left) and net baryon
 number density (right) distribution.}
\end{figure}
Putting the initial time as $\tau=1.0$ fm/$c$, we parameterize the initial
energy density distribution and net baryon number distribution as simple
gaussian forms (Fig.\ \ref{fig:initial}) \cite{Ishii_PRD46}.\footnote{In
Ref.\ \cite{Ishii_PRD46}, gaussian parameterizations are adopted to
temperature and net baryon number distribution.} The parameters
in the model should be chosen so that calculated single-particle spectra
reproduce the experimental results of Au+Au central collisions in the 130
GeV/nucleon at RHIC. The parameter set is summarized in
Table.\ \ref{tbl:param}. Note that we use Bjorken's solution \cite{Bjorken}
$Y_{\text{L}}=\eta$ only as an initial condition for the longitudinal flow
velocity. Initial transverse flow is simply neglected. Once the freeze-out
hypersurface is fixed, one can calculate the single particle spectra via
Cooper-Frye formula \cite{Cooper_Frye}. We assume that the freeze-out process
occurs a specific temperature $T_{\text{f}}$. This condition corresponds to 
a constant energy density in the present calculation because of small
baryonic chemical potential. By virtue of the Lagrangian
hydrodynamics, we expect that contributions of the time-like hypersurface are
small and the space-like hypersurface dominates the particle
emission at freeze-out; we simply put $k^\mu d\sigma_\mu\simeq k^\tau
d\sigma_\tau$.\footnote{It is well
known that the Cooper-Frye formula has an ambiguity in the treatment of the
time-like hypersurface \cite{Sinyukov_ZPHYS43}}
We take hadrons from decay of resonances into account as well as
directly emitted particles from the freeze-out hypersurface. 
We include decay processes $\rho\rightarrow 2\pi$, $\omega\rightarrow 3\pi$,
$\eta\rightarrow 3\pi$, $K^* \rightarrow \pi K$, and $\Delta\rightarrow
N\pi$ \cite{Sollfrank_ZPHYS52,Hirano_PRL}. The resonances
are also assumed to be emitted from the freeze-out hypersurface.
\begin{figure}[ht]
 \includegraphics[width=3.375in]{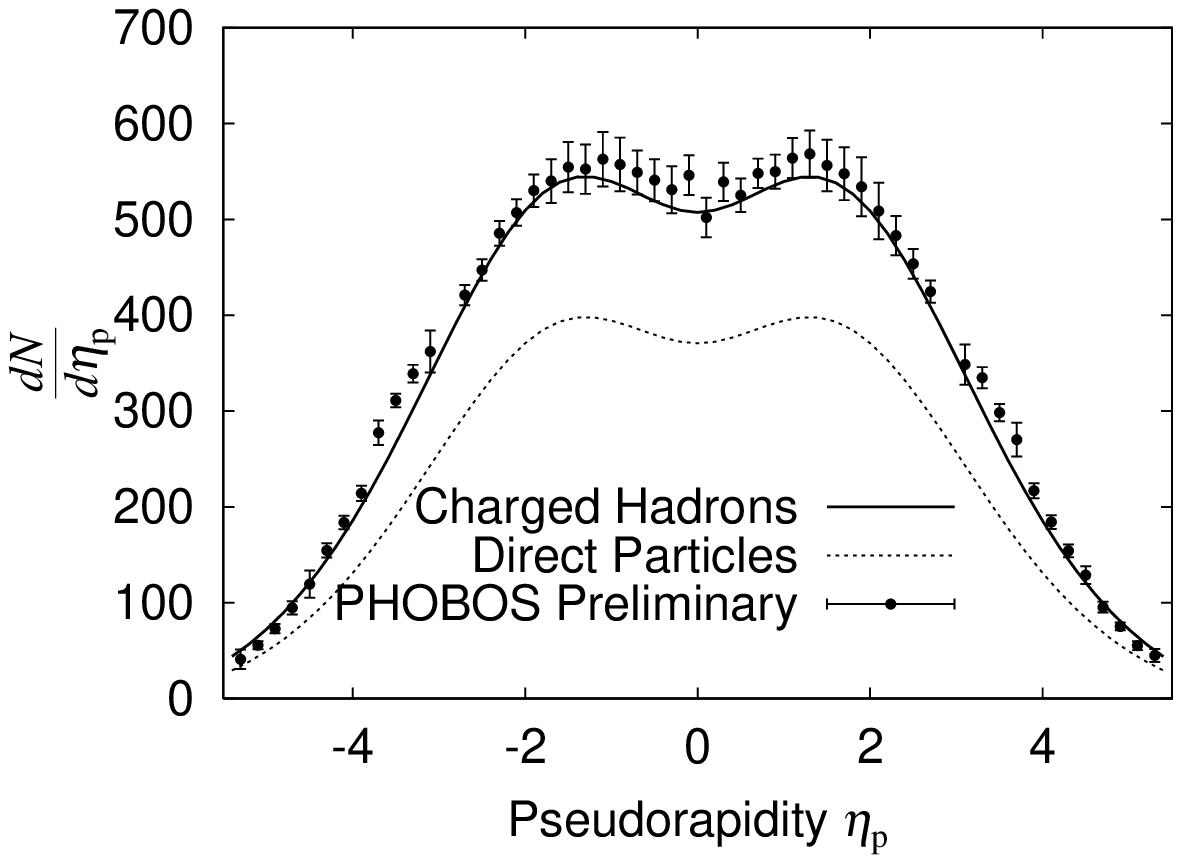}
 \caption{\label{fig:dndeta}Pseudorapidity $\eta_{\text{p}}$
 distribution of charged
 hadrons. Solid line shows our result ($\pi, K, p$) including
 resonance contribution. Dotted line denotes contribution of the directly
 emitted particles from the freeze-out hypersurface. Closed circles are
 preliminary result from the PHOBOS Collaboration \cite{PHOBOS_dndeta}.}
 \includegraphics[width=3.375in]{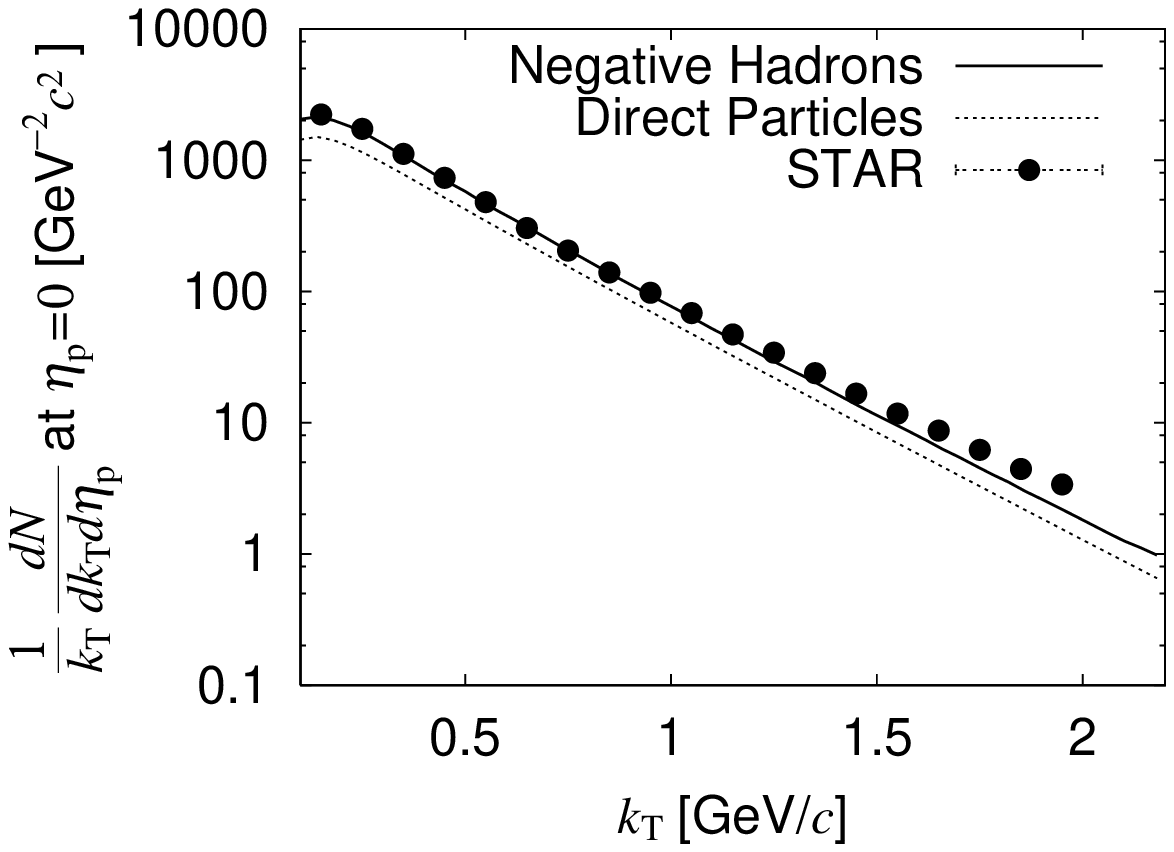}
 \caption{\label{fig:ptspectra}Transverse momentum spectrum of negatively
 charged hadrons. As in Fig.\ \ref{fig:dndeta}, solid line and dotted line
 show total number of particles and directly emitted particles from the
 freeze-out hypersurface, respectively. Closed circles are data from the
 STAR Collaboration \cite{STAR_PRL87}.}
\end{figure}
\begin{figure}[ht]
 \includegraphics[width=3.375in]{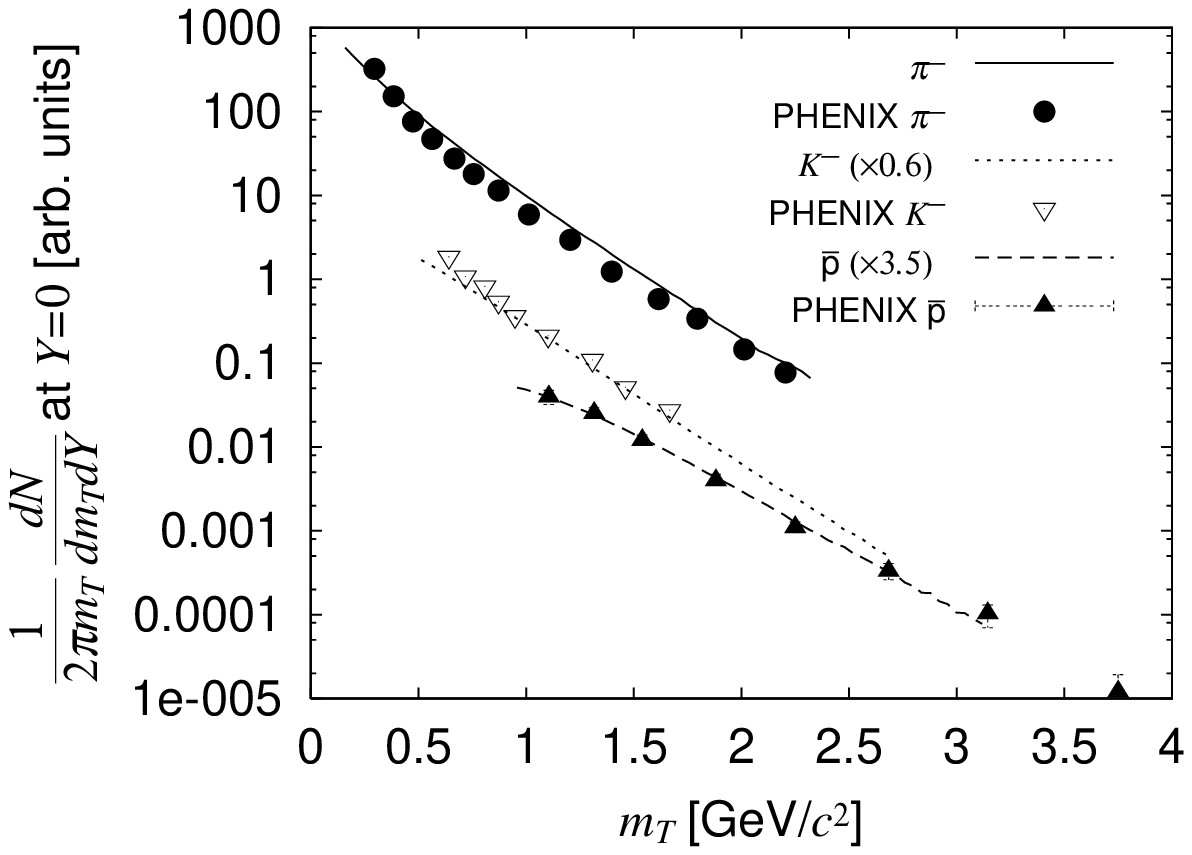}
 \caption{\label{fig:mtspectra}Transverse mass spectra of negatively charged
 hadrons. Solid line, dotted line and dashed line denote $\pi^-$, $K^-$
 and $\bar{p}$ yield of our result. $K^-$ and $\bar{p}$ spectra are scaled
 down by factor 0.1 and 0.01, respectively. Closed circles, open triangles
 and closed triangles are preliminary data from the PHENIX Collaboration
 \cite{PHENIX_mt}. }
 \includegraphics[width=3.375in]{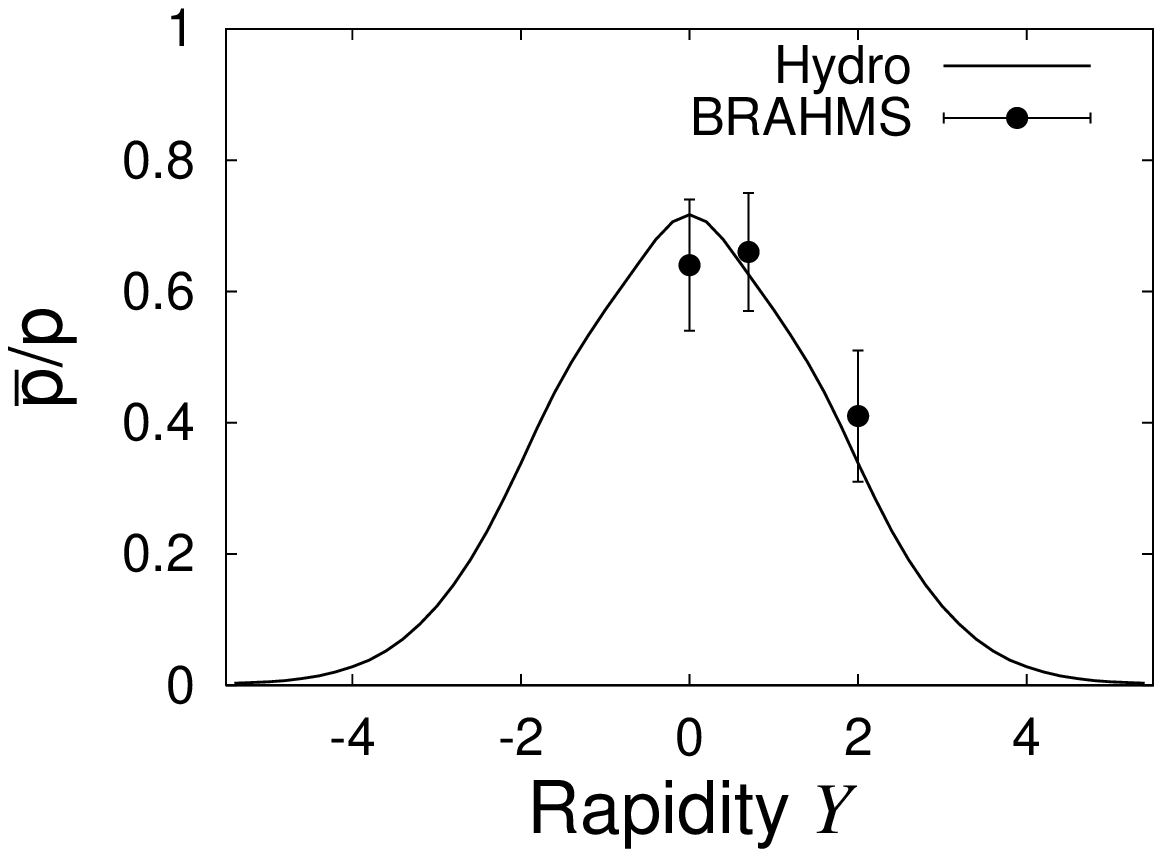}
 \caption{\label{fig:ratio}Rapidity dependence of anti-proton to proton
 ratio. Experimental data are taken from the BRAHMS Collaboration
 \cite{BRAHMS_PRL}.}
\end{figure}
Figure \ref{fig:dndeta} shows pseudorapidity distribution of charged
hadrons. Figure \ref{fig:ptspectra} shows transverse momentum spectrum of
negatively charged hadrons. In both figures, our results well reproduce the
experimental data. We display the transverse mass spectra of identified
negatively charged hadrons ($\pi^-$, $K^-$, $\bar{p}$) in Fig.\
\ref{fig:mtspectra}. All slopes of the spectra are well reproduced by our
calculation. For $K^-$ and $\bar{p}$, our results
fail to reproduce the particle numbers but slopes agree well. In
Fig.\ \ref{fig:ratio}, our result shows excellent agreement with the
experimental data of anti-proton to proton ratio as a function of
rapidity. However, the absolute numbers of anti-protons was not 
enough (Fig.\ \ref{fig:mtspectra}). Though we here assume
the same freeze-out condition for all particle species, the discrepancy may
indicate the more complicated mechanism. 
 \begin{table}
 \caption{\label{tbl:param}Parameter set for the Au+Au collisions.}
 \begin{ruledtabular}
 \begin{tabular}[t]{p{6cm}l}
  Maximum initial energy density  $E_0$ & 6.0 GeV/fm$^3$ \\ 
  Maximum initial net baryon density  $n_{\text{B0}}$ & 0.125 fm$^{-3}$ \\
  Longitudinal gaussian width $\sigma_{\eta}$ of initial energy density &
  1.47 \\
  Longitudinal extension $\eta_0$ of the flat region in the initial energy
  density & 1.0 \\
  Longitudinal gaussian width $\sigma_{\text{D}}$ of the initial net baryon
  density & 1.4 \\ 
  Space-time rapidity $\eta_{\text{D}}$ at maximum of the initial net baryon
  distribution & 3.0 \\
  Gaussian smearing parameter $\sigma_{\text{r}}$ of the transverse profile
  & 1.0 fm \\
  Freeze-out temperature $T_{\text{f}}$ & 125 MeV \\
 \end{tabular}
 \end{ruledtabular}
 \end{table}
 \begin{table}
 \caption{\label{tbl:output}Output.}
 \begin{ruledtabular}
 \begin{tabular}[t]{p{6cm}l}
  Net baryon number & 131 \\
  Mean chemical potential at freeze-out $\langle \mu_{\text{B}} \rangle$ &
  76.1 MeV \\
  Mean transverse flow velocity $\langle v_{\text{T}} \rangle $ of the fluid
  at $|\eta|<0.1$ & 0.509$c$ \\
  Lifetime of the QGP phase $\tau_{\text{QGP}}$ & 2.92 fm/$c$ \\
  Lifetime of the mixed phase $\tau_{\text{MIX}}$ & 12.61 fm/$c$ \\
  Total lifetime of the fluid $\tau_{\text{HAD}}$ & 18.94 fm/$c$ \\
 \end{tabular}
 \end{ruledtabular}
 \end{table}
From the parameter set in Table.\ \ref{tbl:param}, we can see that the energy
density is  sufficient for the QGP production. However, the
energy density, $\epsilon_{\text{max}}=$ 6.0 GeV/fm$^3$ which corresponds to
the temperature of 229 MeV, is only 5\% higher than SPS
\cite{fullpaper}.  Large collision energy at RHIC leads
to very large volume of the hot matter. Longitudinal extension of the hot
matter, $\sigma_\eta+\eta_0=2.47$, is about 2.3 times larger than the one of
SPS \cite{fullpaper}. Initial energy density itself depends strongly on the
initial time $\tau_0$ which corresponds to the spatial size in the
longitudinal direction. Hence, much higher energy density should be obtained
if we assume an earlier thermalization time. Furthermore, we do not include
the thickness of incident nuclei but use almost flat profile with gaussian
smearing for the transverse direction (see Fig.\
\ref{fig:initial}). Therefore, maximum energy density of our model is smaller
than other models which take the thickness into account
\cite{Kolb_PLB500,Zschiesche_nucl7037,Hirano_PRC}. 
The average energy density at $\eta=0$ transverse plane in our model is 3.9
GeV/fm$^3$. Total energy of the fluid of our model corresponds to 25290 GeV,
99 \% of total collision energy. 
We display the outputs from the fluid in Table.\ \ref{tbl:output}. In the
present model, net baryon number is much smaller than total baryon number of
incident nuclei.  $\langle \mu_{\text{B}} \rangle$ means average of the
chemical potential on the whole freeze-out hypersurface. $\langle
v_{\text{T}} \rangle$ is the average transverse velocity of the fluid element
in $|\eta| \leq 0.1$ at the freeze-out. ``Lifetimes'' of the each phase are
also shown in Table.\ \ref{tbl:output}. Space-time evolution of the fluid is
displayed in Fig.\ \ref{fig:evo}. The large area between $T=160$ MeV and
$T=158$ MeV means large space-time volume of the mixed phase. 
\begin{figure}[ht]
 \includegraphics[width=3.375in]{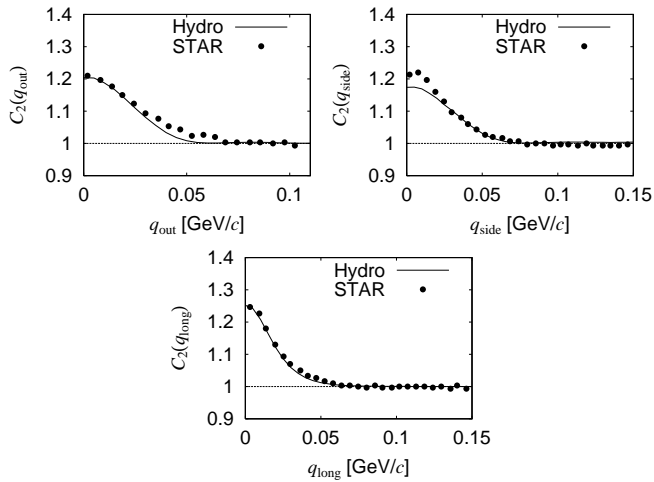}
 \caption{\label{fig:c2}Two-particle Bose-Einstein correlation functions
 for $\pi^-$.  Upper-left, upper-right and lower figures show outward,
 sideward and longitudinal correlation functions, respectively. In each
 figure, correlation function is integrated with respect to other two
 components from 0 to 35 MeV/$c$. Experimental data (closed circles) are taken
 from the STAR Collaboration \cite{STAR_HBT}.}
\end{figure}
\begin{figure}[ht]
 \includegraphics[width=3.375in]{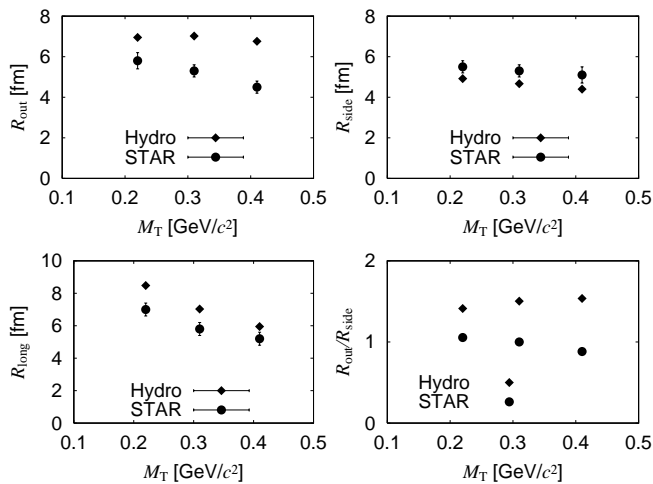}
 \caption{\label{fig:radii}$M_{\text{T}}$ dependence of $\pi^-$ HBT radii.
 Upper-left, upper-right, lower-left and lower-right figures correspond to
 outward, sideward, longitudinal HBT radii and ratio of outward HBT radii to
 sideward HBT radii, respectively. Experimental data (closed circles)
 are taken from the STAR Collaboration \cite{STAR_HBT}.}
\end{figure}

The two-particle correlation function for chaotic source is calculated
through
\begin{equation}
 C_2(q^\mu, K^\mu)=1+\frac{|I(q^\mu, K^\mu)|^2}{I(0,k_1^\mu)I(0,k_2^\mu)}
\end{equation}
where $K^\mu=(k_1^\mu+k_2^\mu)/2$,  $q^\mu = k_1^\mu-k_2^\mu$,
respectively \cite{Shuryak_PLB44, Hama_PRD37}. We put
\begin{equation}
 I(q^\mu, K^\mu)= \int K_\mu d\sigma^\mu(x) \sqrt{f(k_1,x)f(k_2,x)}
 \, e^{iq_\nu x^\nu}, \label{eq:c2}
\end{equation}
so that $I(0,k^\mu)$ reduces to the Cooper-Frye formula with $f(k,x)$ being
the Bose-Einstein distribution function. 
As a first trial, we neglect the contributions from
resonance decay for simplicity. Figure \ref{fig:c2} shows the projected
correlation functions for $\pi^-$ with $|Y| \leq 0.5$ and $0.125 <
K_{\text{T}} < 0.225$
GeV/$c$. In each correlation function, our results are corrected by
a common $\lambda$ factor whose value is 0.6. The other origin of the
reduction is integration with respect to other components of the relative
momenta over the range $0 < q_i < 35$ MeV/$c$. Despite the neglection 
of resonance contribution,\footnote{It has
been discussed that the hadronic cloud significantly affects the HBT radii
\cite{Soff_PRL}.} both outward and longitudinal correlation
functions show good agreements with the experiment. 
Finally, we compare pair transverse mass
$M_{\text{T}}=\sqrt{K_{\text{T}}^2+m^2}$
dependence of the HBT radii extracted through the Gaussian fit to the
three-dimensional correlation function (Fig.\ \ref{fig:radii}). 
We obtain almost consistent sideward HBT radii but obtained outward and
longitudinal HBT radii are larger than experimental data. This tendency
is common to the all hydrodynamical calculation \cite{Zschiesche_nucl7037,
Heinz_hep0111075}. However, the difference between our results and data, 1 fm
in HBT radii, corresponds to only 5 MeV in the correlation function (Fig.\
\ref{fig:c2}). As for $R_{\text{out}}$, a naive interpretation of the
experimental result is that high $M_{\text{T}}$ pions come from high
temperature source and the experimental results indicate
short lifetime of the high temperature region. On the other hand, if the
freeze-out picture works well, i.e., most of particles are emitted from the 
3-dimensional hypersurface, the experimental data suggest very short
freeze-out duration or strong opaqueness of the source
\cite{Heiselberg_EPJC1}. Though our model naturally shows opaque property due
to hydrodynamical flow \cite{Morita_PRC}, the freeze-out time duration,
$\Delta t = \sqrt{\langle t^2 \rangle - \langle t \rangle^2}$ where $\langle
A(x) \rangle = \int d^4 x A(x)S(x,K) / \int d^4x S(x,K)$ with $S(x,K)$ being
the source function \cite{Chapman_PRC}, of our
model is about 5 $\sim$ 7 fm/$c$ which is longer than the SPS case
\cite{Morita_PRC, fullpaper}. This long time duration of our model leads to
the large $R_{\text{out}}/R_{\text{side}}$ \cite{Rischke_NPA608}.

In summary, we present a hydrodynamical-model calculation for the 130
GeV/nucleon Au+Au collisions data from the RHIC experiments. The present
numerical solution indicates that the produced quark-gluon plasma in the RHIC
has much lower baryon density, slightly higher energy density, and several
times larger extension in the longitudinal direction than in the SPS case, if
we compare at the same initial time, $\tau_0=1$ fm/$c$.
We present the first analysis of the pion Bose-Einstein correlation data at
RHIC based on the hydrodynamical model which takes into account both
longitudinal and transverse expansion appropriately. The result is partly
consistent with the experimental result but obtained $R_{\text{out}}$ and
$R_{\text{long}}$ are slightly larger than the experimental
data. In this work, we concentrate our discussion on the RHIC data. More
detailed discussion including the comparison with the SPS data will be given
in the forthcoming paper \cite{fullpaper}.

\begin{acknowledgments}
 The authors are much indebted to Professor I. Ohba and Professor H. Nakazato
 for their fruitful comments. They also would like to thank Dr. H. Nakamura
 and other members of high energy physics group at Waseda University for
 helpful discussions. One of the authors (C.N.) would like to acknowledge the
 financial support by the Soryushi Shogakukai. This work is partially
 supported by Waseda University Grant for Special Research Projects
 No.2001A-888 and Waseda University Media Network Center.
\end{acknowledgments}

\end{document}